\documentclass[twocolumn,showpacs,preprintnumbers,amsmath,amssymb,pre]{revtex4}

\usepackage{graphicx}
\usepackage{dcolumn}
\usepackage{bm}
\usepackage{enumerate}
\usepackage{amsmath}
\usepackage{color}

\begin{document}

\title{Subdiffusion equation with Caputo fractional derivative with respect to another function in modelling diffusion in a complex system consisting of matrix and channels}

\author{Tadeusz Koszto{\l}owicz}
 \email{tadeusz.kosztolowicz@ujk.edu.pl}
 \affiliation{Institute of Physics, Jan Kochanowski University,\\
         Uniwersytecka 7, 25-406 Kielce, Poland}

\author{Aldona Dutkiewicz}
 \email{szukala@amu.edu.pl}
 \affiliation{Faculty of Mathematics and Computer Science,\\
Adam Mickiewicz University, Uniwersytetu Pozna\'nskiego 4, 61-614 Pozna\'n, Poland}

\author{Katarzyna D. Lewandowska}
 \email{katarzyna.lewandowska@gumed.edu.pl}
 \affiliation{Department of Radiological Informatics and Statistics, \\Medical University of Gda\'nsk,\\ Tuwima 15, 80-210 Gda\'nsk, Poland}
				
\author{S{\l}awomir W\c{a}sik}
 \email{s.wasik@ujk.edu.pl}
 \affiliation{Department of Medical Physics and Biophysics, Institute of Physics, Jan Kochanowski University,\\
         Uniwersytecka 7, 25-406 Kielce, Poland}

\author{Micha{\l} Arabski}
 \email{arabski@ujk.edu.pl}
 \affiliation{Institute of Biology, Jan Kochanowski University,\\ Uniwersytecka 7, 25-406 Kielce, Poland}

\date{\today}

\begin{abstract}

Anomalous diffusion of antibiotic (colistin) in a system consisting of packed gel (alginate) beads immersed in water is studied experimentally and theoretically. The experimental studies are performed using the interferometric method of measuring concentration profiles of diffusing substance. We use the $g$--subdiffusion equation with the fractional Caputo time derivative with respect to another function $g$ to describe the process. The function $g$ and relevant parameters define anomalous diffusion. We show that experimentally measured time evolution of the amount of antibiotic released from the system determine the function $g$. The process can be interpreted as subdiffusion in which subdiffusion parameter (exponent) $\alpha$ decreases over time. The $g$--subdiffusion equation, which is more general than the "ordinary" fractional subdiffusion equation, can be widely used in various fields of science to model diffusion in a system in which parameters, and even a type of diffusion, evolve over time. We postulate that diffusion in a system composed of channels and matrix can be described by the $g$--subdiffusion equation, just like diffusion in a system of packed gel beads placed in water.

\end{abstract}

\maketitle

\section{Introduction\label{sec1}} 

Subdiffusion is a process in which the movement of molecules is very hindered by a complex internal structure of the medium  \cite{mk,ks,barkai2000,barkai2001,cherstvy,chechkin,denisov,mjcb}. This process was observed, among others, in diffusion of sugars in agarose gel \cite{kdm} and in antibiotics diffusion in a bacterial biofilm \cite{km,kmwa}, the reference list can be extended significantly. A distinctive feature of ``ordinary'' subdiffusion is the relation $\sigma^2(t)\sim t^\alpha$, where $\sigma^2$ is the Mean Square Displacement of a diffusing particle and $\alpha\in (0,1)$. In general, subdiffusion occur when time of a molecule to jump is anomalously long. Within the Continuous Time Random Walk model, for ``ordinary'' subdiffusion the distribution of a waiting time for a molecule to jump $\psi$ has a heavy tail, $\psi(t)\sim 1/t^{\alpha+1}$, the average value of this time is infinite. ``Ordinary'' subdiffusion can be described by a differential equation with a time derivative of fractional order controlled by the parameter $\alpha$ \cite{mk,ks,barkai2000,barkai2001,cherstvy,chechkin,denisov,mjcb,compte}. 

The situation is more complicated when molecules diffuse in a system consisting of matrix and channels. The channels are defined here as "free spaces" in the matrix which usually have a complicated geometric structure. The channels can be ``free spaces'' in a porous medium, tubules, or spaces between packed beads. Diffusion occurs mainly in channels, but molecules can diffuse into and out of the matrix. Diffusion of various substances in a system consisting of channels and matrix has been considered in medicine, biology, engineering, geology, agriculture, and other fields of science. The examples are diffusive release of vitamin from collagen \cite{teimouri}, nutrients from a fertilizer to water and sand \cite{du}, fertilizers from beads \cite{fertahi,santos}, diffusion of oxygen in soils \cite{neira}, the process of active ingredients release which can be used in reduction of groundwater pesticide pollution \cite{paradelo}, insulin release from chitosan beads \cite{jose}, drug release from alginate beads \cite{iskakov,yuan1} and from coating beads \cite{sawicki}. Such studies are needed to establish the conditions under which the optimal dose of the drug is released \cite{macha,fu}. Theoretical models of the processes mentioned above have been based on diffusion--reaction equations \cite{ray,peppas,siepmann,weiser,liu,neira,fu,mircioiu}, ``ordinary'' subdiffusion equation with Caputo derivative in which source term is involved \cite{chang}, ``ordinary'' subdiffusion-reaction equation \cite{lenzi}, and on scaling approach \cite{oar}. Subiffusion parameters depend on disorder of packing beads \cite{palombo,hlus} and on the structure of beads \cite{tamm}. 

We assume that the process could be described by the ``ordinary'' subdiffusion equation with a fixed parameter $\alpha$ at some initial time interval. The normal diffusion equation may be treated as a special case of the ``ordinary'' subdiffusion equation for $\alpha=1$, thus we do not exclude normal diffusion from further considerations. However, getting the molecules into the beads and back again can slow down subdiffusion. Similar processes in which diffusion of molecules is interrupted by their trapping in an immobile zone have been considered in \cite{dcsm,oar,ling,bmrv}. Traps can be small caves with narrow passages. Such traps change the time scale of the process \cite{hs,hs1}. To describe the process we use a subdiffusion equation with the fractional Caputo derivative with respect to another function $g$ \cite{kd,kd1,tk2022} which controls the slowness of the process. We call this equation $g$--subdiffusion equation. The key is to determine the $g$ function experimentally. For this purpose it is convenient to define a function describing diffusion process and controlled by the function $g$ that is relatively easily measurable experimentally. 

In the following, we experimentally study diffusion of an antibiotic (colistin) in a system in which gel (alginate) beads soaked with the antibiotic are placed in water. We use the $g$--subdiffusion equation to describe the process and show that the time evolution of the amount of the antibiotic released from the system allows the determination of the function $g$. 

\section{Experiment\label{sec2}}

\begin{figure}[htb]
\centering{%
\includegraphics[scale=0.55]{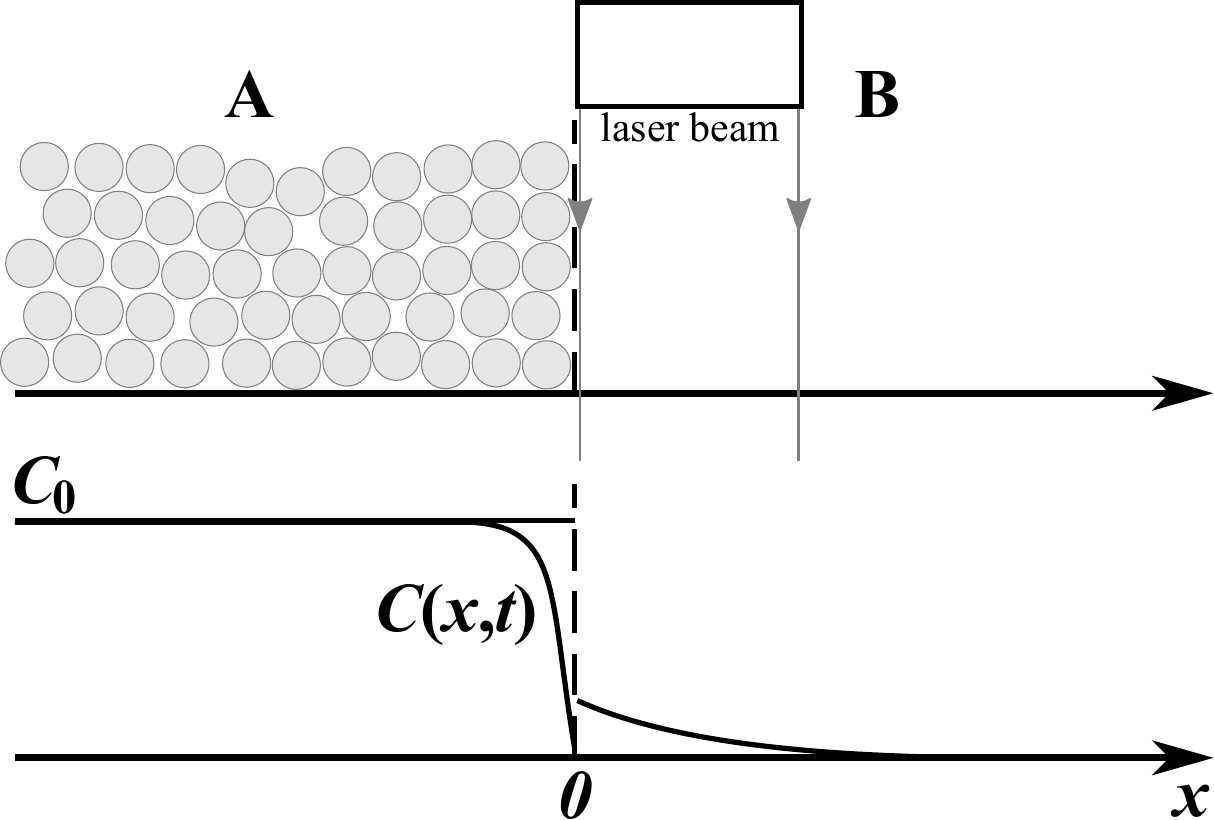}}
\caption{Top panel: scheme of the system used in the experiment, region (vessel) $A$ contains alginate beads impregnated with colistin placed in water and region $B$ contains water at the initial moment. As the beads are non--transparent to the laser beam, interferometric measurement of concentration is only possible in region $B$. Lower panel: $C$ denotes the concentration of the antibiotic, $C_0$ is the initial concentration in region $A$.}
\label{fig1}
\end{figure}

The system used for the experimental study consists of two regions $A\;(x<0)$ and $B\;(x>0)$ separated by a thin membrane located at $x=0$, see Fig. \ref{fig1}. We assume that the system is homogeneous in the plane perpendicular to the $x$-axis, so it is effectively one-dimensional. The membrane, which is very permeable to diffusing particles, keeps the beads in region $A$. At the initial moment, all antibiotic is in beads, it is distributed uniformly in each bead. The antibiotic concentration is measured by means of the interferometric method. Since the region containing the beads is not transparent to the laser beam, measurements of the antibiotic concentration can only be made in region $B$. We focus our attention on a time evolution of the amount of the antibiotic $N(t)$ that has diffused from region $A$ to $B$.

The scheme of the system is shown in Fig. \ref{fig1}. The system consists of two vessels separated by a thin membrane. The vessel sizes are: the cross-sectional area is $S=7\times 10^{-5}\;{\rm m^2}$, the length of the vessel $A$ (measured along the $x$ axis) is $L_A=10^{-2}\;{\rm m}$, and the length of the vessel $B$ is $L_B=5\times 10^{-2}\;{\rm m}$. Antibiotic soaked alginate beads were made as follows. $1\;{\rm mg}$ of colistin in the form of methanesulfonate sodium (Fluka, Germany) was dissolved in $1\;{\rm ml}$ of $1.5\%$ alginate solution. Using an automatic pipette, the alginate solution with an antibiotic was added dropwise to a calcium chloride solution of $0.15\; {\rm mol/m^3}$ concentration. As a hydrogel carrier the sodium alginate (SAFC, USA) was used. Due to the biocompatibility as a crosslinking agent calcium cations were applied, the source of which was calcium chloride (POCH S.A., Poland). A single bead has a volume of $15\;{\rm \mu l}$. $27$ beads were in the vessel $A$, the ratio of the total volume of all beads to the volume of the vessel $A$ is equal to $0.58$. Within the homogeneous medium approximation the initial colistin concentration in the vessel $A$ was calculated using the formula $\mathcal{C}_0=\mathcal{N}_0/V_A$, where $\mathcal{N}_0$ is the total amount of colistin in all beads at the initial moment. We obtained $\mathcal{C}_0=0.50\;{\rm mol/m^3}$. Linear initial concentration used in theoretical model $C_0=S\mathcal{C}_0$ is $C_0=3.5\times 10^{-5}\;{\rm mol/m}$. Concentration profiles of antibiotic were determined in region $B$ for different times $t\in [120\;{\rm s},2400\;{\rm s}]$.  

For experimental study we used colistin in the form of methanesulfonate sodium (Fluka, Germany). Antibiotic solutions were prepared in double deionized water. The hydrogel carrier was sodium alginate (SAFC, USA), and calcium cations were selected as the cross-linking agent due to their biocompatibility. The source of calcium cations was calcium chloride (POCH S.A., Poland).  
In order to prepare alginate beads with colistin, $1\;{\rm mg}$ of colistin was dissolved in $1\;{\rm ml}$ of a $1.5\%$ alginate solution and mixed thoroughly. Then, using an automatic pipette, the alginate-antibiotic solution was dripped into the $0.15\;{\rm mol/m^3}$ calcium chloride solution also containing the antibiotic. As a result of gelation, beads having a volume approximately $15\;{\rm \mu l}$ were obtained. 

The colistin concentration has been measured by means of the laser interferometric method \cite{kdm,kmwa,wad,kwl}.  
\begin{figure}
  \centering
  \includegraphics[scale=0.43]{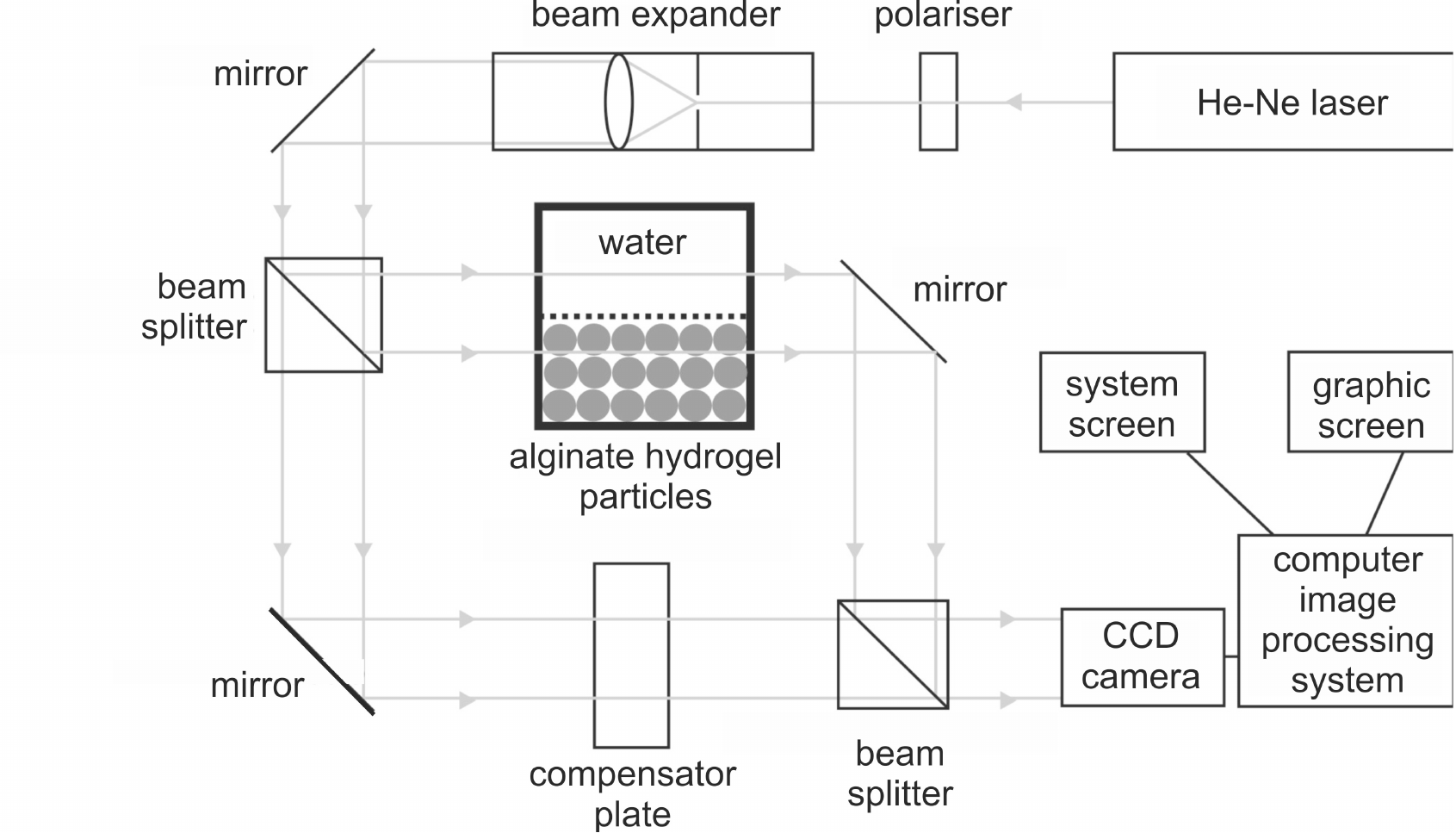}
  \caption{Experimental setup. Detailed description is in the text.}
  \label{aparatura}
\end{figure}
Fig.~\ref{aparatura} presents the sketch of measuring apparatus. The main element of the apparatus is the double-beam Mach-Zehnder interferometer with a laser illumination system and a computerized system for recording and processing of interference images. The laser beam produced by a $15\;{\rm mW}$ He–Ne laser is spatially filtered and next, using the beam expander, is transformed into a parallel beam of width of $80\;{\rm mm}$ and later is split into two beams. One of them passes through both cuvettes parallel to the membrane surface, while the other being a reference beam goes directly through the compensation plate to the light detecting system where it superimposes with the laser beam which passing through the diffusion cell. The result is the formation of interference fringes, see Fig.~\ref{interfer}. 
\begin{figure}
  \centering
  \includegraphics[scale=0.07]{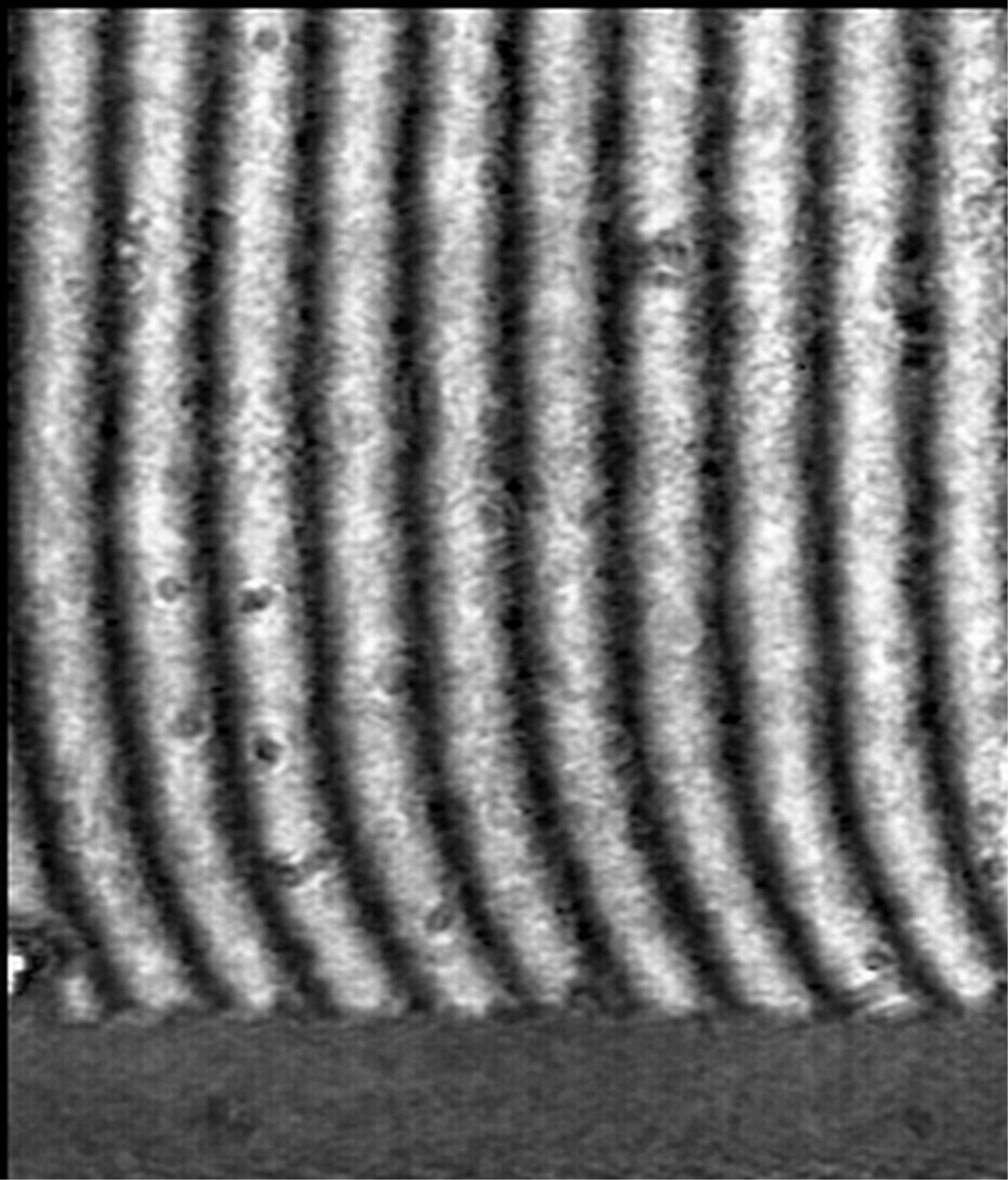}
  \caption{Interferogram obtained after $20\;{\rm min}$.}
  \label{interfer}
\end{figure}

\begin{figure}
  \centering
  \includegraphics[scale=0.25]{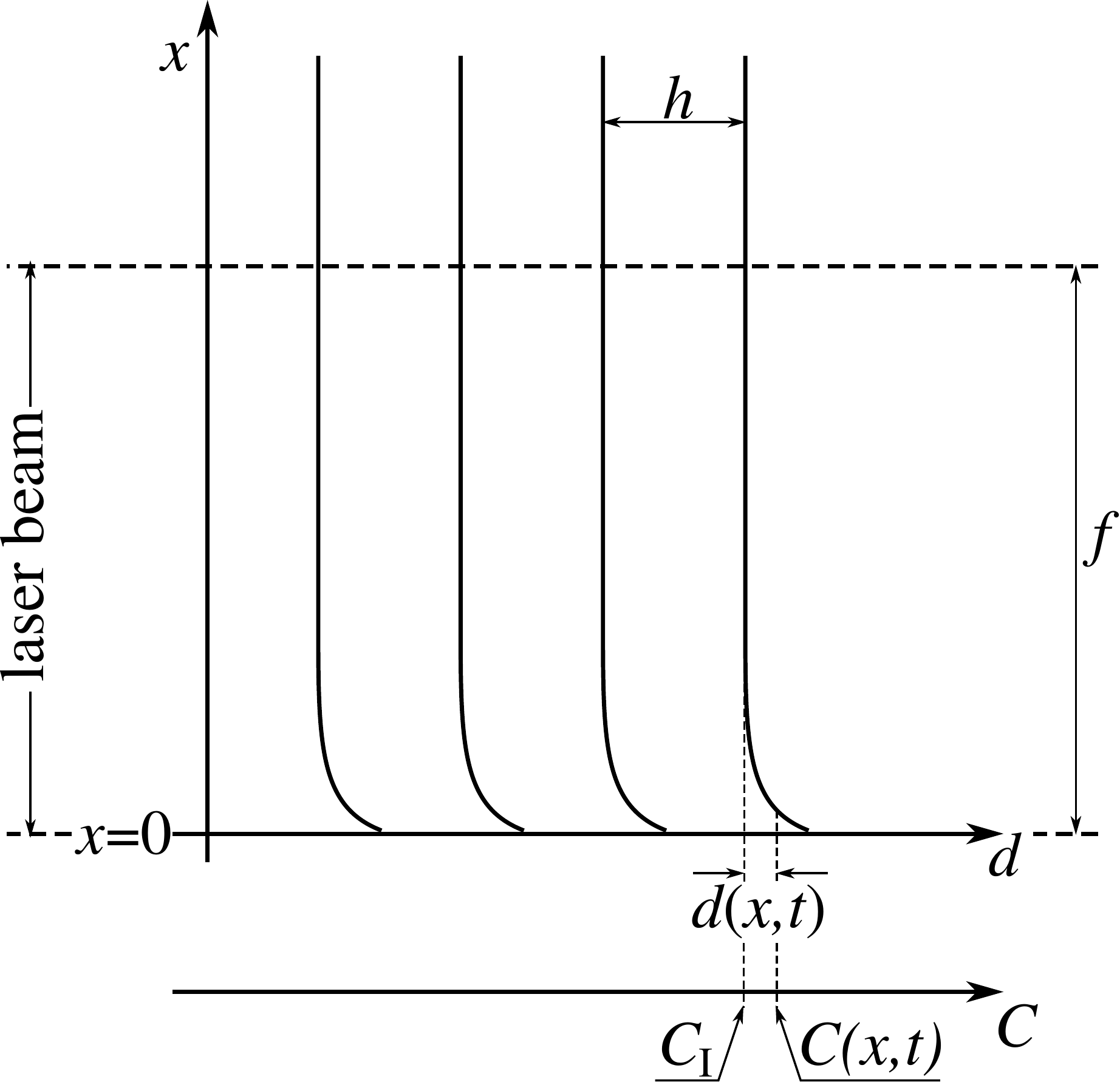}
  \caption{Schematic drawing of the interference fringes.}
  \label{prazki}
\end{figure}

The course of the interference fringes is determined by the refractive index of the solution, which in turn depends on its concentration. Interference fringes are straight when the solute is homogeneous and bend when a concentration gradient is non--zero. The magnitude of the deviation of the interference fringe at a given point with respect to a undisturbed fringe $d(x,t)$ (see Fig.~\ref{prazki}) reflects the changes in the refractive index between these points and thus provides information about changes in the concentration of the substance between the points. 

The relation between the change of the substance concentration $\Delta C(x,t)$ and the change of the refractive index $\Delta n(x,t)$ is linear \cite{yen,torres,boyu,yuan}, 
\begin{equation}\label{st}
  \Delta C(x,t)=a\Delta n(x,t)\;,
\end{equation}
where $\Delta C(x,t)=C_{\rm I}-C(x,t)$, $C_{\rm I}$ is the initial concentration, 
\begin{equation}\label{wsp_a}
  \Delta n(x,t)=\frac{\lambda d(x,t)}{hf}\;,
\end{equation}
where $\lambda$ denotes a wavelength of the laser light ($632.8\;{\rm nm}$), $h$ is the distance between the interferometric fringes in the area where they are parallel and $f$ denotes thickness of solution layer along the course of the laser beam. The coefficient $a$ was determined in a separate experiment using the interference refractometer (Zeiss). In the range of tested concentration the dependence of refractive index on the concentration of colistin solution is linear and the value of the parameter $a$ is $2.92\times10^3\;{\rm mol/m^3}$. The sign of the deviation of the interference fringe $d$ depends on the measurement setup settings. In the experiment, $d(x,t)<0$ means that in the observed region the interference fringe deviates to the right and the substance flows in from other regions what means that the concentration in this region is increasing relative to initial concentration. When $d(x,t)>0$, the interference fringe deviates to the left and the substance flows away to the other regions what means the concentration in this region is decreasing relative to initial concentration. 

The diffusion cell consists of two vessels made of glass with a very high optical homogeneity separated by a horizontally located thin membrane (see Fig.~\ref{aparatura}). Initially, the alginate beads with colistin of concentration $1\;{\rm mg/ml}$ were placed close together completely filling the volume of lower cuvette, while the upper cuvette was filled with pure water, thus $C_{\rm I}=0$. Then, an antibiotic diffuses to the upper cuvette. The duration of the experiment was $40\;{\rm min}$ and interferograms were recorded with time interval equals $2\;{\rm min}$. The measurements was conducted under isothermal conditions at a temperature of $T=295\pm0.3\;{\rm K}$.

\section{Theory\label{sec3}} 

We use the $g$-subdiffusion equation to model antibiotic diffusion in a system consisting of antibiotic soaked alignian beads densely packed in water. Normal diffusion or subdiffusion may occur in the spaces between the beads. As we have mentioned, the motivation behind the use of the $g$-subdiffusion equation is that the release of the antibiotic from the gel beads, as well as the possibility of the antibiotic particles re-penetrating the beads, can slow down antibiotic diffusion in the system. 

Diffusion in region $A$ is a combination of two processes: subdiffusion of antibiotic molecules inside the bead to their exit outside and their further diffusion mainly in a ``free space'' between the beads. The border between a bead and the free space is not an obstacle for the particles exiting the bead. However, it is an obstacle for the particles trying to re-enter the bead because then the particle must hit a pore on the bead surface. We suppose that the return of the antibiotic molecule to a bead is possible when the molecule moves in region $A$ and has frequent contact with the bead surfaces. When a molecule diffuses in region $B$ near the border with region $A$ the molecule contact with the beads is much less frequent and the molecule is unlikely to return to region $A$. We assume that the boundary between the regions can be regarded as an absorbing wall for molecules located in region $A$. The most often used boundary condition at the absorbing wall is zero concentration of diffusing substance at the wall. The amount of antibiotic that is in region $B$ is equal to the amount of the antibiotic that left region $A$ at the same time. Therefore, in the following we consider diffusion of the antibiotic in region $A$. We use the approximation of a homogeneous medium for region $A$ and assume that the medium structure does not change in time. Then, the medium consisting of beads and ``free spaces'' between beads has assigned subdiffusion parameters such as for a homogeneous medium; similar approximation has been used in modeling of diffusion in disordered system of spheres \cite{palombo,hlus}. We also assume that the parameters are independent of time and a spatial variable and the antibiotic is distributed homogeneously in region $A$ at the initial moment. The parameters presented later in this paper and Eq. (\ref{eq4}) show that $\sigma\approx 0.036\;{\rm cm}$ for $t=2400\;{\rm s}$ whereas the length of vessel $A$ is $1.0\;{\rm cm}$. Thus, we suppose that the influence of the outer wall of the vessel $A$ on molecules diffusing from $A$ to $B$ is negligibly small; at the outer wall the concentration is still $C_0$. To simplify the calculations, we assume that the wall is located at $-\infty$. Diffusion of antibiotic molecules in the vessel $A$ is described by the $g$--subdiffusion equation Eq. (\ref{eq2}), the initial condition is $C(x,0)=C_0$, the boundary conditions are $C(-\infty,t)=C_0$ and $C(0^-,t)=0$. 

Subdiffusion is often described by the subdiffusion equation with the Riemann--Liouville fractional time derivative \cite{mk,ks,barkai2000,barkai2001,compte}, which may be converted to the following form 
\begin{equation}\label{eq01}
\frac{^C\partial^\alpha \tilde{C}(x,t)}{\partial t^\alpha}=D\frac{\partial^2\tilde{C}(x,t)}{\partial x^2}
\end{equation}
with the ``ordinary'' Caputo derivative of the order $\alpha\in (0,1)$ defined as 
\begin{equation}\label{eq02}
\frac{^C\partial^\alpha f(t)}{\partial t^\alpha}=\frac{1}{\Gamma(1-\alpha)}\int_0^t(t-u)^{-\alpha}f'(u)du,
\end{equation}
$D$ is a subdiffusion coefficient given in the units of ${\rm m^2/s^\alpha}$. We call Eq. (\ref{eq01}) the ``ordinary'' subdiffusion equation, its solution is denoted here as $\tilde{C}$. Recently, differential equations with a fractional Caputo derivative with respect to another function $g$ (the $g$--Caputo fractional derivative) have been considered \cite{fahad,almeida,jarad,jarad1}. For $\alpha\in (0,1)$ this derivative is defined as
\begin{equation}\label{eq1}
\frac{^Cd^{\alpha}_g f(t)}{dt^\alpha}=\frac{1}{\Gamma(1-\alpha)}\int_0^t (g(t)-g(t'))^{-\alpha} f'(t')dt',
\end{equation}
the function $g$, which is given in a time unit, fulfils the conditions $g(0)=0$, $g(\infty)=\infty$, and $g'(t)>0$ for $t>0$. 
Involving this derivative in the diffusion equation we get the $g$--subdiffusion equation
\begin{equation}\label{eq2}
\frac{^C \partial^{\alpha}_g C(x,t)}{\partial t^\alpha}=D\frac{\partial^2 C(x,t)}{\partial x^2}.
\end{equation}
When $g(t)\equiv t$ we have ``ordinary'' subdiffusion equation.
Solutions $\tilde{C}$ of the ``ordinary'' subdiffusion equation and $C$ of the $g$--subdiffusion equation are related to each other as follows \cite{kd,kd1}
\begin{equation}\label{eq3}
C(x,t)=\tilde{C}(x,g(t)), 
\end{equation}
if the boundary conditions, the initial condition, and the parameters $\alpha$ and $D$ are the same for both equations. Thus, the $g$--subdiffusion equation describes the subdiffusion process with changed time variable. If diffusion is described by Eq. (\ref{eq2}), then random walk of a single molecule is characterized by the relation \cite{kd}
\begin{equation}\label{eq4}
\sigma^2(t)=\frac{2Dg^\alpha(t)}{\Gamma(1+\alpha)}.
\end{equation}
Choosing the function $g$ appropriately, we can obtain different functions $\sigma^2$ that have been derived from other models, e.g. $g\sim {\rm log}^\alpha t$ for ultraslow diffusion and $g\sim t^{2/d_w}$ for diffusion on a fractal, $d_w$ is the fractal dimension of a medium; an overview of $\sigma^2$ derived from different models is presented in \cite{mjcb,csm}.

The key is to find the $g$ function on the basis of experimental results. A function providing the function $g$ is a time evolution of the total amount of the substance that has diffused from region $A$ to $B$, 
\begin{equation}\label{eq013}
N(t)=\int_{-\infty}^0 [C_0-C(x,t)]dx, 
\end{equation}
see Fig. \ref{fig1}. 

Solutions to the $g$--subdiffusion equation can be obtained by means of the $g$-Laplace transform method. This transform is defined as 
\begin{equation}\label{eq09}
\mathcal{L}_g[f(t)](s)=\int_0^\infty {\rm e}^{-s g(t)}f(t)g'(t)dt, 
\end{equation}
it has the following property \cite{fahad,jarad}
\begin{equation}\label{eq010}
\mathcal{L}_g\left[\frac{^Cd^{\alpha}_g f(t)}{dt^\alpha}\right](s)=s^\alpha\mathcal{L}_g[f(t)](s)-s^{\alpha-1}f(0) 
\end{equation}
that makes the procedure for solving Eq. (\ref{eq2}) similar to the procedure for solving ``ordinary'' subdiffusion equation Eq. (\ref{eq01}) using the ``ordinary'' Laplace transform $\mathcal{L}[f(t)](s)=\int_0^\infty {\rm exp}(-st)f(t)dt$ \cite{kd}. In terms of the $g$--Laplace transform the $g$--subdiffusion equation is 
\begin{equation}\label{eq011}
s^\alpha\mathcal{L}_g[C(x,t)](s)-s^{\alpha-1}C_0=D\frac{\partial^2\mathcal{L}_g[C(x,t)](s)}{\partial x^2}.
\end{equation}
The solution to this equation for the boundary conditions $\mathcal{L}_g[C(-\infty,t)](s)=C_0/s$, $\mathcal{L}_g[C(0^-,t)](s)=0$, and the initial condition $C(x,0)=C_0$ is
\begin{equation}\label{eq5}
\mathcal{L}_g[C(x,t)](s)=\frac{C_0}{s}\left(1-{\rm e}^{\sqrt{s^\alpha/D}x}\right).
\end{equation}
The $\mathcal{L}_g$ transform is related to the ``ordinary'' Laplace transform by the formula 
\begin{equation}\label{eq012}
\mathcal{L}_g[f(t)](s)=\mathcal{L}[f(g^{-1}(t))](s),
\end{equation} 
this formula is helpful in calculating the inverse $g$--Laplace transforms \cite{fahad,jarad}. From the relation $\mathcal{L}_g[1](s)=1/s$, the $g$--Laplace transform of Eq. (\ref{eq013}), and Eq. (\ref{eq5}) we obtain $\mathcal{L}_g[N(t)](s)=C_0\sqrt{D}/s^{1+\alpha/2}$. Using the formula $\mathcal{L}_g^{-1}[1/s^{1+\beta}](t)=g^\beta (t)/\Gamma(1+\beta)$, $\beta,s>0$, we get 
\begin{equation}\label{eq6}
N(t)=\kappa g^{\alpha/2}(t),
\end{equation}
where 
\begin{equation}\label{eq05}
\kappa=\frac{C_0\sqrt{D}}{\Gamma(1+\alpha/2)}. 
\end{equation}
For ``ordinary'' subdiffusion we have
\begin{equation}\label{eq7}
N(t)=\kappa t^{\alpha/2}.
\end{equation}
The function for normal diffusion is obtained from the above equations puting $\alpha=1$,
\begin{equation}\label{eq04}
N(t)=\frac{2C_0\sqrt{D}}{\sqrt{\pi}}\sqrt{t}.
\end{equation}

\section{Determination of the function $g$ and the parameters $\alpha$ and $D$\label{sec4}} 

\begin{figure}[htb]
\centering{%
\includegraphics[scale=0.31]{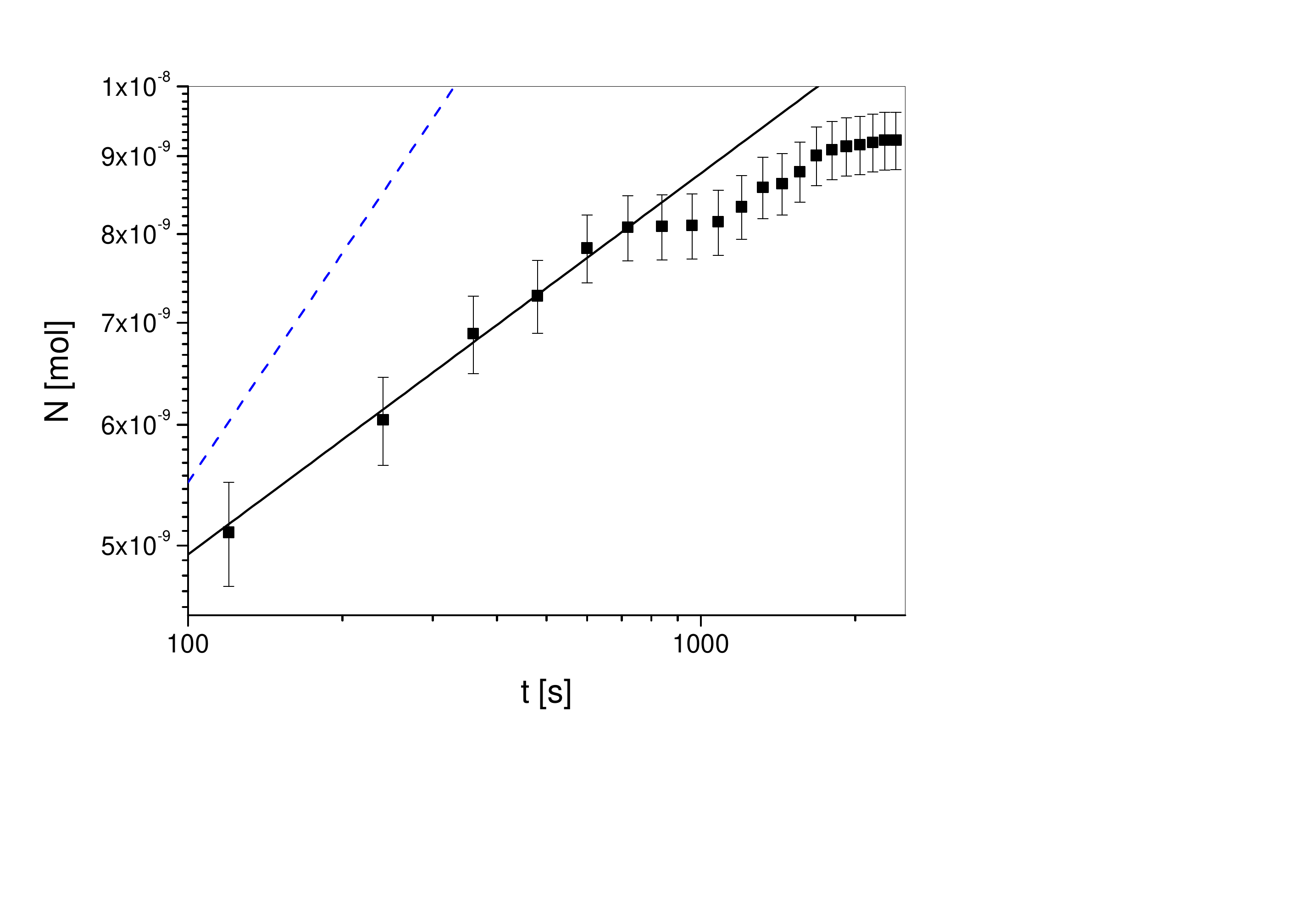}}
\caption{Plots of the power functions $N(t)=1.57\times 10^{-9} t^{0.25}$ for ``ordinary'' subdiffusion (solid line) and $N(t)=0.55\times 10^{-9} \sqrt{t}$ for normal diffusion (dashed line) in the {\rm log--log} scale, the empirical results are denoted by symbols. The function for normal diffusion is an example, it does not represent any process considered in this paper.}
\label{fig5}
\end{figure}

We assume that properties of the medium do not change rapidly, thus the function $N$ is smooth, i.e. its derivative is continuous. Since $N(t)\sim \sigma(t)$, we use $N$ to identify the type of diffusion. When $N$ is a power function with an exponent less than $1/2$ the process can be treated as subdiffusion.
The empirical results and the power functions $N$ Eq. (\ref{eq7}) with $\alpha=0.5$ (solid line) and Eq. (\ref{eq04}) for normal diffusion (dashed line) are presented in the logarithmic scale in Fig. \ref{fig5}. The slope of the lines representing $N$ depends on the parameter $\alpha$ only. 

From Fig. \ref{fig5} we conclude that: (1) the process under study is not normal diffusion, (2) the process is not ``ordinary'' subdiffusion with constant parameter $\alpha$ over the entire time domain, (3) the process can be treated as ``ordinary'' subdiffusion only in some initial time interval (we estimate it as $t\leq 480\;{\rm s}$, which corresponds to four initial points in the plot), (4) for longer times the process is slower than ``ordinary'' subdiffusion mentioned in point (3). 

Based on the above conclusions, we assume that: (i) for $t\in[0,480\;{\rm s}]$ the process is ``ordinary'' subdiffusion with parameters $\alpha$ and $D$, (ii) for later times the process is $g$--subdiffusion with function $g$ which generates the relation
\begin{equation}\label{eq06}
N(t)=\kappa t^{\tilde{\alpha}(t)/2},
\end{equation}
where $\tilde{\alpha}(t)$ fulfils the conditions $0<\tilde{\alpha}(t)\leq 1$ and $\tilde{\alpha}(0)=\alpha$. Eqs. (\ref{eq6}) and (\ref{eq06}) provide $g(t)=t^{\tilde{\alpha}(t)/\alpha}$. Parameters $\alpha$ and $D$ in the $g$--subdiffusion equation Eq. (\ref{eq2}) are the same as for the initial ``ordinary'' subdiffusion process. 
Eq. (\ref{eq06}) is a generalization of Eq. (\ref{eq7}) for the case of a time-varying subdiffusion parameter (exponent).

In further considerations we assume
\begin{equation}\label{eq07}
\tilde{\alpha}(t)=\frac{\alpha}{1+\beta t},
\end{equation}
where $\beta$ is a parameter measured in the unit of $1/{\rm s}$.
Comparing Eqs. (\ref{eq6}) and (\ref{eq06}), the latter with the exponent Eq. (\ref{eq07}), we obtain
\begin{equation}\label{eq08}
g(t)=\kappa^{2/\alpha}t^{1/(1+\beta t)}.
\end{equation}

\begin{figure}[htb]
\centering{%
\includegraphics[scale=0.31]{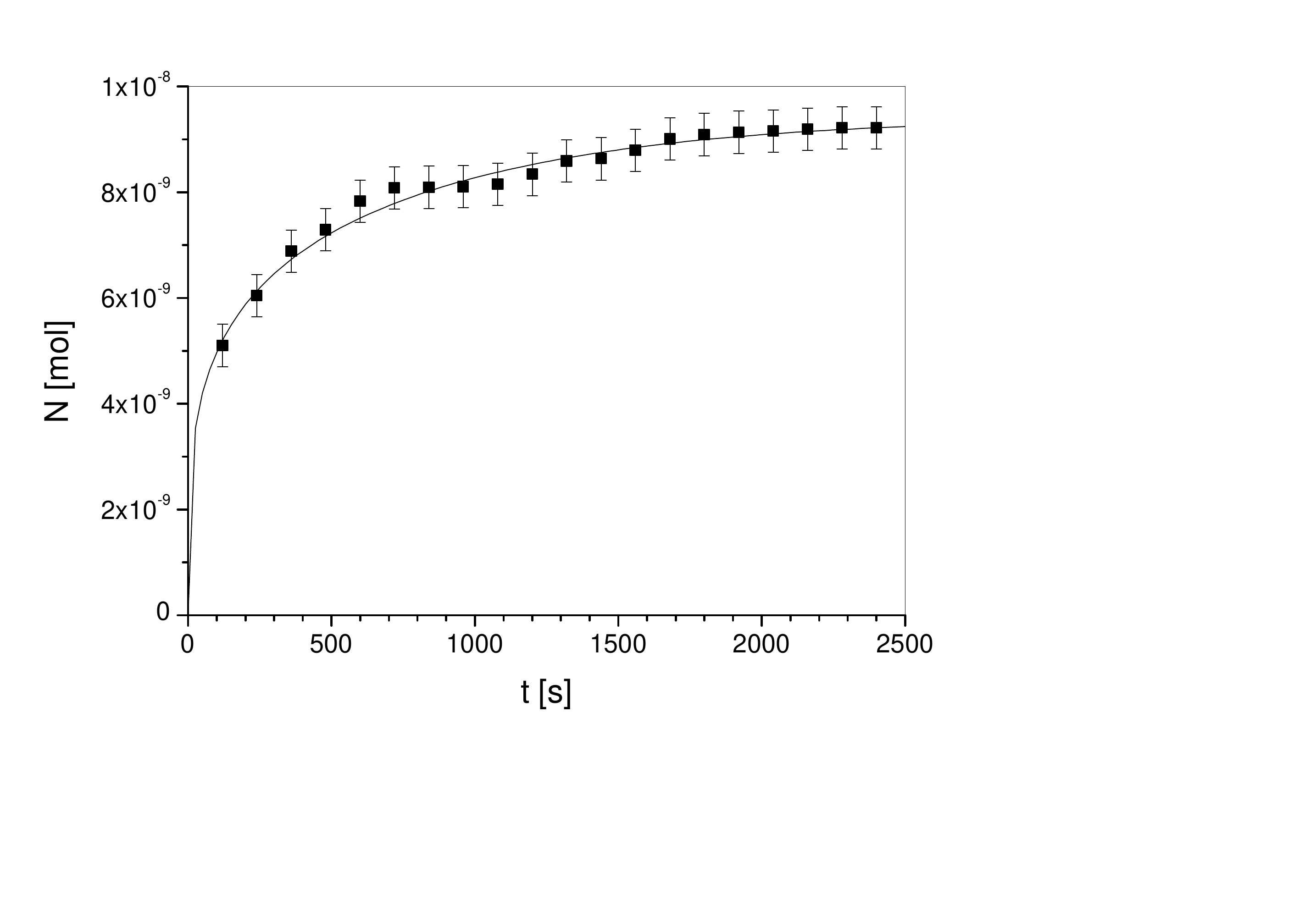}}
\caption{Plot of the function $N$ Eq. (\ref{eq06}) with $\tilde{\alpha}(t)$ given by Eq. (\ref{eq07}) (solid line) for $\kappa=1.57\times 10^{-9}\;{\rm mol/s^{0.25}}$, $\alpha=0.5$, and $\beta=4.3\times 10^{-5}\;{\rm 1/s}$, the empirical results are denoted by symbols.}
\label{fig6}
\end{figure}

Fig. \ref{fig5} shows that for $t\in[120\;{\rm s},480\;{\rm s}]$ the data are described well by the function $N$ Eq. (\ref{eq7}) with $\alpha=0.5$ and $\kappa=1.57\times 10^{-9}\;{\rm mol/s^{0.25}}$. From the last equation, the value of initial concentration $C_0$, and Eq. (\ref{eq05}) we get  $D=0.58\times 10^{-9}\;{\rm m^2/s^{0.5}}$. Knowing $\kappa$ and $\alpha$, we fit the function $N$ Eq. (\ref{eq06}) with the exponent given by Eq. (\ref{eq07}) to the empirical data in the whole time domain, the fit parameter is $\beta$ only. As shown in Fig. \ref{fig6}, for $\beta=4.3\times 10^{-5}\;1/{\rm s}$ the fit is good.

\section{Final remarks\label{sec5}}

We postulate: {\it Diffusion in a system composed of channels and matrix can be described by the $g$--subdiffusion equation Eq. (\ref{eq2})}, 
just like diffusion in a system of packed gel beads placed in water. The solution to the $g$--subdiffusion equation provides the time evolution of the amount of colistin released from region $A$ which is consistent with the empirical results. We mention that different models of normal and anomalous diffusion, describing the release of substances from the (sub)diffusive medium, give power functions $N(t)\sim t^\beta$ (when ``ordinary'' subdiffusion occurs in $A$) or exponential functions of the form $N(t)\sim 1-a{\rm e}^{-bt^\beta}$ in the long time limit \cite{peppas,teimouri,yuan,siepmann,mircioiu,tk2017,tk2019}. In our study, the $g$--subdiffusion equation provides $N$ as a power function with exponent evolving over time. Based on Fig. \ref{fig5} we conclude that it is not possible to model the diffusion process in region $A$ over the entire time domain using the "ordinary" subdiffusion equation with a fixed parameter $\alpha$. 

The fact that the experimentally obtained function $N$ is well approximated by a power function with decreasing subdiffusion parameter $\tilde{\alpha}(t)$ we interpret as follows. Colistin (Polymyxin E) is a cationic antimicrobial peptide. Due to the presence of positively charged five L-diaminobutyric acid (L-Dab) amino groups in colistin structure, it is possible that this molecule interacts with calcium alginate. Two mechanisms of polycations binding to alginate have been proposed: an electrostatic interaction and the formation of a calcium alginate gel by displacement of calcium ions in the presence of polycations \cite{thu}. Taking into account the interaction of colistin with alginate and the complex geometric structure of the channels in the alginate gel, subdiffusion of colistin in alginate beads is expected.
The first stage of the process is the subdiffusive release of molecules located in the beads close to their surfaces to the space between the beads. This process is ruled by distribution of waiting time to take particle next step with a heavy tail controlled by the parameter $\alpha$, as for ``ordinary'' subdiffusion. Initially, releasing antibiotic molecules from region $A$ to $B$ is subdiffusion of molecules from beads located at the border between the regions. Later, the process may change its nature when molecules located in $A$ in the layers more distant from the border diffuse into region $B$. These molecules have a more complicated path from the inside of vessel $A$ to the vessel $B$. 

The $g$--subdiffusion equation offers greater possibilities for modeling subdiffusion processes compared to the ``ordinary'' subdiffusion equation. Recently, anomalous diffusion equations with various fractional derivatives have been considered, see the references cited in Ref. \cite{kd}. Unfortunately, such equations often do not have a stochastic interpretation. The interpretation of Eq. (\ref{eq2}) is based on Eq. (\ref{eq3}). We mention that the derivation of Eq. (\ref{eq2}) can be based on the "ordinary" Continuous Time Random Walk model \cite{mk,ks}, in which, as in Eq. (\ref{eq3}), the time variable in the distribution of waiting time for a molecule to jump is changed, $\psi(t)\rightarrow\psi(g(t))$ \cite{kd1}.

The model of diffusion in a matrix with channels system can be used, among others, in the mathematical characteristic of drug delivery system for wound healing \cite{johnson}, the extracellular matrix mimetic to deliver and retain therapeutic cells at the site of administration for tissue engineering and regenerative medicine \cite{prestwich,hussey} or ``smart'' hydrogels as thermo- or pH-responsive \cite{ferreira}. Alginate and the other polysaccharides are the most commonly used materials for polymyxin delivery systems as provide suitable drug loading efficiency and controlled drug release. So, it seems to be important to characteristic these effects on a theoretical level for proper design of the chemical structure of gels as a controlled matrix for drug releasing. It is crucial for effective bacteria eradication by a lethal concentration of antibiotics obtain in the wound healing environment or stimuli the tissue regeneration \cite{dub}.

\end{document}